\input{aipcheck}

\documentclass[
  ] {aipproc}

 \layoutstyle{8x11single}

\newcommand{\text}{}

\newcommand{\Ne}{N^{obs-e}} 
\newcommand{\Nt}{N^{obs-t}}
\newcommand{\Me}{M^{obs-e}}
\newcommand{\Mt}{M^{obs-t}}
\renewcommand{\tt}{t_t}
\newcommand{\te}{t_e}

\newcommand{\ct}{c_t}
\newcommand{\ce}{c_e}
\newcommand{\lampt}{{lamp-t}}
\newcommand{\lampe}{{lamp-e}}

\newcommand{\ft}{f^{obs-t}}
\newcommand{\fe}{f^{obs-e}}
\newcommand{\vt}{v_t}
\newcommand{\ve}{v_e}

\newcommand{\obst}{observer-train}
\newcommand{\obse}{observer-embankment}

\newcommand{\Obse}{Observer-embankment}
\newcommand{\Oe}{O_{emb}}
\newcommand{\Ot}{O_{train}}
\newcommand{\BEQ}{\begin{eqnarray}}
\newcommand{\EEQ}{\end{eqnarray}}
\newcommand{\BEA}{\begin{eqnarray}}
\newcommand{\EEA}{\end{eqnarray}}

\begin{document}

\title{Observational Derivation of  Einstein's \\``Law of the Constancy of the Velocity of Light {\it in Vacuo}"}

\author{Claudia Pombo}{address={Amsterdam, the Netherlands}}

\author{Theo M. Nieuwenhuizen}
{address={Institute for Theoretical Physics, 
Valckenierstraat 65, 1018 XE Amsterdam, The Netherlands}}

\date{today}

\begin{abstract}
On the basis of Galilean invariance and the Doppler formula, combined with 
an observational condition, it is shown that the constancy of the velocity of light {\it in vacuo} 
can be derived, together with time-dilatation and Lorentz contraction. 
It is not necessary to take the constancy as a postulate.

\end{abstract}
\maketitle

\section{Introduction}

In this centenary of Einstein's {\it annus mirabilis}, it is proper to reconsider one 
of his breakthroughs. We shall focus here on his 1905 paper on special relativity \cite{1},
 in the language of the train-embankment setup he discussed in later works \cite{EinsteinTrain}.

Motivated by the results of H. A. Lorentz on electrodynamics \cite{2,3}, involving the invariance 
of the velocity of light, Einstein postulated that the speed of light is the same in all  
non-accelerated systems of reference.
He deduced from this statement, combined with the principle of relativity, the kinematic base 
of the theory of special 
relativity. 

In particular, Einstein drew attention to the
issue of the non-absolutism of time, the effect of time dilatation: a material event lasting a 
certain time for an observer at rest with respect to it, 
lasts a longer time for an observer moving with constant speed with respect to it.
This aspect explains why cosmic muons can be observed on the surface of the earth, 
since their rather short decay time is dilated for observers on the earth, so that part 
of them survive the travel through the earth's atmosphere.

It is known that non-absolutism of time is behind the constancy of the velocity of the light.
But, by itself, this is a difficult notion. In observational terms 
the constancy of the speed of light is also difficult to understand. We do not enter 
here any deep discussion about non-absolutism of time, except that we call attention
to the fact that atomic clocks also emit light and, consequently, they are not so different 
from lamps. 

It is the purpose of the present paper to offer an alternative, pedagogic derivation 
of the constancy of the speed of light. We do not make any assumption on the 
velocity of the light but on numbers of cycles and on frequencies.

In section II we present our setup, in section III we pose our conditions and in
section IV we analyze the situation. We end up with concluding that
our conditions imply a derivation of the constancy of the speed of light. 

\section{The setup of train, embankment, lamps and observers}

A train having speed $v_e>0$ passes an embankment. A point in the train is called
the origin $\Ot$ and at time $t=0$ this passes  the origin of the embankment 
called $\Oe$.
In the setup there are lamps. The first is at the origin of the embankment $\Oe$, 
the second at the origin of the train $\Ot$.
There are also two observers. One uses the point $\Ot$ as reference and is called \obst.
Another one takes the embankment as reference and is called \obse.
We also consider another point on the embankment at a distance $X_e>0$ 
in the direction of the movement of the train, at which \obse \ 
can measure light.

The lamp at the embankment has, according to \obse, a certain frequency $\fe_\lampe$.
The lamp in the train has a different frequency, $\ft_\lampt$ according to \obst.
This frequency is fixed such that, according to \obst, it has the same frequency 
as the light he observes from the lamp at the embankment, 
$\ft_\lampt=\ft_\lampe$.

Both observers have a clock.

\section{Conditions}

\subsection*{ 1: Galilean invariance}

The train moves with speed $v_e>0$ according to the \obse.
We assume that the \obst \ observes that the origin of the embankment 
$\Oe$ moves with respect to him with speed $\vt=-\ve$.

\subsection*{ 2: Applicability of Doppler formula}

The lamp of the embankment produces in time-interval $(0,t_e)$ a number of cycles 
$\Ne_\lampe(t_e)=\fe_\lampe t_e$.
The number $\Me_\lampe(\te)$ of cycles reaching the train in this interval is assumed,
by \obse,  to be
related by the classical Doppler formula  to the number $\Ne_\lampe(\te)$ emitted at the embankment, 
\BEQ \label{Doppler} \Me_\lampe(\te)=(1-\beta_e) \Ne_\lampe(\te), \EEQ
where 
\BEQ \beta_e=\frac{\ve}{\ce}, \EEQ
with $\ce$ the speed of light according to \obse.

\subsection*{3:  Identification of times}
Both observers count $t=0$ for the event where $\Ot$ passes $\Oe$.
Let us denote the time of \obse \ as $\te$ and the time of 
\obst \ as $\tt$. Later we shall see that these times are indeed different.
We relate their times by equating 
the number of cycles observed by \obst \ with the number of cycles presumed by \obse:

\BEQ \Mt_\lampe(\tt)=\Me_\lampe(\te).
\EEQ

\subsection*{4: Conservation of information} 

This is a physical condition: Observer-embankment measures the 
same frequency from lamp-embankment and lamp-train:

\BEQ \fe_\lampt=\fe_\lampe. \EEQ 
 If it is not satisfied, the speed of light will not be constant.

\section{Analysis of frequencies}

\subsection{The frequency for \obst}

The \obst \ counts the number $\Mt_\lampe$ in a period he denotes as  $(0,\tt)$. 
\Obse \ knows about this counting and calls the period $(0,\te)$. 
We shall see that it is necessary to assume that
for \obse \  this counting time is dilated to be 
\BEQ \label{te=gamtt} \te=\gamma_e \tt,\EEQ 
where $\gamma_e$ is for now some unknown factor.
If it would appear to be equal to unity, there would be no time-dilatation. 
But we shall show below that it is equal the Lorentz factor $1/\sqrt{1-\beta_e^2}$.

The frequency of the lamp at the embankment is for \obst \
\BEQ \ft_\lampe=\frac{\Mt_\lampe(\tt)}{\tt}. \EEQ
This becomes equal to
\BEQ \ft_\lampe=\gamma_e (1-\beta_e)\,\frac{\Ne_\lampe(\te)}{\te}=
\gamma_e (1-\beta_e)\,\fe_\lampe. \EEQ
From the expression for $\gamma_e$ that we shall deduce, this will appear to be red-shifted.

In the setup outlined above, the lamp in the train emits light
with exactly this frequency, according to \obst.

In accordance with Eq. (\ref{te=gamtt}), the time to reach the position $X_e$ of 
\obse \ is according to \obst

\BEQ\label{txt=txe}  t_{x;t}=\frac{1}{\gamma_e}\,t_{x;e} \EEQ

\subsection{Frequency of the lamp in the train according to \obse}

Now we repeat this argument for the lamp in the train, which is observed by \obse.
This means that there is an interchange in the role of lamps and clocks.
Whereas the lamp in the embankment was going away from this observer,
there is a time interval, according to \obst \  $(0,t_{x;t})$, 
in which the train approaches position $X_e$. 
Observer-embankment calls this the interval $(0,t_{x;e})$, with  $t_{x;e}=X_e/\ve$.

Consider \obst's time interval $(\tt',t_{x,t})$. In this interval the number of 
cycles produced by the lamp in the train is
\BEQ \Nt_\lampt=\ft_\lampt\,\Delta \tt',\qquad \Delta\tt'=t_{x;t}-\tt', \EEQ
 where the primes indicates that now lamp-train is discussed.
According to the Doppler formula, \obst \ assumes that the number of cycles received by
 \obse \ in the related time-interval is 
\BEQ \Me_\lampt(\Delta\te')=(1-\beta_t) \Nt_\lampt(\Delta\tt'),\EEQ 
where 
\BEQ \beta_t=-\frac{v_t}{\ct}, \EEQ
involves the {\it speed of light as observed by \obst}.
The sign changes with respect to Eq. (\ref{Doppler}) because now the lamp,
located in the train, is approaching $X_e$.

\Obse \ counts these pulses in a time interval $\Delta\te'=t_{x;e}-\te'$. In analogy with previous
case, we have to assume that \obst \ considers this period to last a time
\BEQ \Delta\tt'={\gamma_t}\Delta\te', \EEQ
where also the factor $\gamma_t$ is to be determined.

We now have 
\BEQ \fe_{\lampt}=\frac{\Me_\lampt}{\Delta\te}=\gamma_t(1-\beta_t)\frac{\Nt_\lampt}{\Delta\tt}
 =\gamma_t(1-\beta_t)\ft_\lampt. \EEQ

Because in our setup is designed such that $\ft_\lampt=\ft_\lampe$, it follows from Eq. (7) that

\BEQ \fe_{\lampt}=\gamma_t(1-\beta_t)\gamma_e (1-\beta_e) \fe_\lampe. \EEQ

According to our condition 4, these frequencies are the same, so we conclude that

\BEQ\label{helpme} \gamma_t(1-\beta_t)\gamma_e (1-\beta_e)=1. \EEQ

\subsection{A mirror of the setup}

In our setup, the origin of the train $\Ot$ \ had just passed the origin of the embankment $\Oe$
and moved on.
In an alternative setup, we might have considered the approach of the train,
while \obse \ performs a measurement at the point $-X_e<0$. 
For \obst,  approaching the lamp of the embankment, its frequency would be blue-shifted rather than 
red-shifted, and the lamp in the train is supposed to be adjusted to this new value.

Repeating all steps one-by-one in the two related time-intervals, 
we would deduce that now Eq. (\ref{helpme}) holds with the signs of $\beta_e$ and $\beta_t$ altered,
because of the interchange of approaching and separating. 
Keeping our previous definitions of $\beta_e$ and $\beta_t$, this means that 
 
\BEQ \label{please} \gamma_e(1+\beta_e)\gamma_t (1+\beta_t)=1. \EEQ

\subsection{Derivation of the constancy of the speed of light}

When dividing Eq. (\ref{helpme}) and Eq. (\ref{please}) we obtain
\BEQ \frac{(1-\beta_t)(1-\beta_e)}{(1+\beta_e) (1+\beta_t)}=1. \EEQ
This has the solution
\BEQ \beta_e=-\beta_t\qquad {\rm or} \qquad \frac{\ve}{\ce}=-\frac{\vt}{\ct}. \EEQ

Since by our Galilean condition we have assumed that $\vt=-\ve$, we {\it deduce}
that the speed of light is the
same in both systems of reference, even though they move with respect to each other,

\BEQ  \ce=\ct. \EEQ

Now going back to (\ref{helpme}), we are left with

\BEQ {\gamma_e\gamma_t}=\frac{1}{{1-\beta_e^2}} \EEQ
This reveals that the classical choice $\gamma_e=\gamma_t=1$ is excluded, as is
the choice $\gamma_t=1/\gamma_e$, that one might naively have guessed when comparing (5) and (12). 
Instead, the solution is to assume that both observers are
completely equivalent, so that $\gamma_e=\gamma_t$, which yields
\BEQ \gamma_e=\frac{1}{\sqrt{1-\beta_e^2}}. \EEQ
This is indeed the Lorentz factor. So given our Condition 4, we did have to assume 
time-dilatation in order to solve the problem.

The distance between the origin $\Oe$ and the position $X_e$ of \obse \ is for \obst
\BEQ X_t=-\vt t_{x;t}=\ve\, \frac{t_{x;e}}{\gamma_e}=\frac{1}{\gamma_e}X_e \EEQ
where we used (\ref{txt=txe}). So also the Lorentz contraction is derived 
in this analysis.

\section{Discussion}

We have demanded a few observational  conditions.
The most important one is the condition on conservation of information:
 the frequency directly observed from a lamp
is the same as when this light is absorbed and re-emitted in a moving reference frame.
From this we were able to derive the constancy of the speed of light
{\it in vacuo}, and, with it, the time dilatation and the Lorentz contraction. 
This provides conditions under which the theory of special relativity is valid.

It is interesting to see that in this approach the observers are to a large extent classical:
they adopt the Galilean principle and the Doppler formula applied to the
number of cycles of the light. It is only when their times are compared, that a 
time-dilatation has to be taken into account.
Indeed,  without it our conditions have been shown to lead to contradictions.

The time-dilatation needs not be known to 
the observers themselves, they need not communicate with each other about their findings.
It is only known to ``us'', as external observers, trying to unify the
observations of these two individuals.

The physical presence of observers can as usual be replaced by e.g. 
automated photo camera's,  as is often done in practice.
But somewhere down the line someone is needed to read and reconcile these observations.
This reconciliation is accomplished with Einstein's theory of special relativity, which holds provided
 the conditions for our analysis are valid.

\section*{Acknowledgments}
C.P. acknowledges hospitality at the University of V\"axj\"o, where
part of this work was done, as well as  partial support by the `Stichting 
voor Fundamenteel Onderzoek der Materie' (FOM), which is financially
supported by the `Nederlandse Organisatie voor Wetenschappelijk
Onderzoek (NWO)'.

\end{document}